\documentstyle[emulateapj,psfig]{article}

\makeatletter

\newenvironment{inlinefigure}{%
\def\@captype{figure}%
\noindent\begin{minipage}{0.999\linewidth}\begin{center}}
{\end{center}\end{minipage}\smallskip}
\makeatother

\lefthead{Barger et al.}

\begin{document}
\title{Very High Redshift X-ray Selected AGN
in the {\it Chandra} Deep Field-North}
\author{
A.~J.~Barger,$\!$\altaffilmark{1,2,3}
L.~L.~Cowie,$\!$\altaffilmark{3}
P.~Capak,$\!$\altaffilmark{3}
D.~M.~Alexander,$\!$\altaffilmark{4}
F.~E.~Bauer,$\!$\altaffilmark{4}
W.~N.~Brandt,$\!$\altaffilmark{4}
G.~P.~Garmire,$\!$\altaffilmark{4}
A.~E.~Hornschemeier$\!$\altaffilmark{5,6}
}

\altaffiltext{1}{Department of Astronomy, University of Wisconsin-Madison,
475 North Charter Street, Madison, WI 53706}
\altaffiltext{2}{Department of Physics and Astronomy,
University of Hawaii, 2505 Correa Road, Honolulu, HI 96822}
\altaffiltext{3}{Institute for Astronomy, University of Hawaii,
2680 Woodlawn Drive, Honolulu, HI 96822}
\altaffiltext{4}{Department of Astronomy \& Astrophysics,
525 Davey Laboratory, The Pennsylvania State University,
University Park, PA 16802}
\altaffiltext{5}{Chandra Fellow}
\altaffiltext{6}{Department of Physics and Astronomy, Johns
Hopkins University, 3400 North Charles Street, Baltimore, MD
21218}

\slugcomment{To appear in The Astrophysical Journal Letters}

\begin{abstract}
Deep {\it Chandra} X-ray exposures provide an
efficient route for locating optically faint
active galactic nuclei (AGN) at high redshifts.
We use deep multicolor optical data to search
for $z>5$ AGN in the
2~Ms X-ray exposure of the {\it Chandra} Deep Field-North.
Of the 423 X-ray sources 
bright enough ($z'<25.2$) for a color-color analysis,
at most one lies at $z=5-6$ and none at $z>6$. 
The $z>5$ object is
spectroscopically confirmed at $z=5.19$. 
Only 31 of the 77 sources with $z'>25.2$
are undetected in the 
$B$ or $V$ bands at the $2\sigma$ level
and could lie at $z>5$. There are too few moderate luminosity 
AGN at $z=5-6.5$ to ionize the intergalactic medium.
\end{abstract}


\section{Introduction}
\label{secintro}

Most of our current information on very high redshift 
active galactic nuclei (AGN) relates to the very high 
luminosity tail of the luminosity function (LF). 
At $z>5$ the Sloan Digital Sky Survey (SDSS) probes to 
an $1450$~\AA\ absolute magnitude of about $-26$ 
(\markcite{fan01a}Fan et al.\ 2001a, b),
much brighter than the average luminosity of
AGN at lower redshifts (\markcite{pei95}Pei 1995). 
Calculations of the contributions of high-redshift AGN to 
the ionizing flux in the Universe must 
therefore rely on large extrapolations
(e.g., \markcite{madau99}Madau, Haardt, \& Rees 1999),
and deeper observations of the LF are needed
to constrain this quantity directly.  
Such observations can also test
models of supermassive black hole formation
(e.g., \markcite{haiman98}Haiman \& Loeb 1998, 1999; 
\markcite{haehnelt98}Haehnelt, Natarajan, \& Rees 1998).

We can efficiently search for faint, high-redshift AGN 
using combined ultradeep X-ray and optical imaging. 
Since at high redshifts observed-frame X-ray bands 
correspond to very high rest-frame energies, we can obtain
a relatively complete sample of AGN, including sources 
surrounded by very high column densities of gas and 
dust. Some Compton-thick sources may be omitted 
from the sample
but would not be expected to contribute much to the ionizing light.
As there are only a handful of stars 
in X-ray samples, there is little 
problem in separating red stars from high-redshift AGN.
Here we use deep broad-band optical imaging and
follow-up optical spectroscopy to measure the $z>5$ AGN
population in the {\it Chandra} Deep Field-North (CDF-N)
2~Ms exposure.
We assume $\Omega_M=1/3$, $\Omega_\Lambda=2/3$ and
$H_o=65$~km~s$^{-1}$~Mpc$^{-1}$.

\section{Data}
\label{secobs}

The deepest X-ray image is the $\approx2$~Ms
CDF-N exposure centered on the Hubble Deep Field-North. 
D. M. Alexander et al.\ (in preparation)
merged samples detected in seven X-ray 
bands to form a catalog of 503  
point sources over an area of 460~arcmin$^2$.
 Near the aim 
point the limiting fluxes are 
$\approx 1.5\times 10^{-17}$~erg~cm$^{-2}$~s$^{-1}$ 
($0.5-2$~keV) and $\approx 10^{-16}$~erg~cm$^{-2}$~s$^{-1}$ ($2-8$~keV). 
The sensitivity degrades with off-axis angle beyond
$\sim 6'$, rising to about
$10^{-15}$~erg~cm$^{-2}$~s$^{-1}$
($0.5-2$~keV) and $10^{-14}$~erg~cm$^{-2}$~s$^{-1}$
($2-8$~keV) at the maximum angle of $\sim 14'$.
The $2-8$~keV number counts and completeness levels in
the 1~Ms exposure of the CDF-N are described in 
\markcite{cowie02a}Cowie et al.\ (2002a). 

Johnson $B$ and $V$, 
Cousins $R$ and $I$, and Sloan $z'$ images were obtained 
on the Subaru\footnote{The Subaru telescope is operated by the
National Astronomical Observatory of Japan.}
8.2~m telescope's
Prime Focus Camera (Suprime-Cam;  
\markcite{miya98}Miyazaki et al.\ 2002). 
The observations and reductions are summarized in 
P. Capak et al. (in preparation). The exposure
times are $z'$ (3.9 hrs), $I$ (2.9 hrs), $R$ (5.3 hrs),
$V$ (8.4 hrs), and $B$ (1.7 hrs). The image
quality is $\approx 0.8''$, except in
the $V$ ($1.0''$) and $R$ ($1.1''$) bands.  
Magnitudes and images are given in
\markcite{barger02}Barger et al.\ (2002; 1~Ms sample) and
A. J. Barger et al. (in preparation; 2~Ms).
We chose the optical position to be
the highest surface brightness peak in the $z'$ image within $1.5''$ 
of the X-ray position.
For most sources the optical
counterpart is unambiguous, but in some cases (about $7\%$),  
there is more than one possible counterpart
(see, e.g., Fig.~2 of \markcite{barger02}Barger et al.\ 2002). 
We discuss alternative counterparts in
\S~\ref{optfaint}.
For off-axis sources in extended galaxies, and 
optically faint sources, we used
the X-ray position.

We measured magnitudes  
in $2''$ diameter apertures on
images smoothed to the lowest resolution image ($R$)
to provide the most accurate colors.
We estimated total magnitudes by correcting
the $2''$ diameter magnitudes to $6''$ diameter 
using a median offset ($\sim0.8$ for each filter) measured 
on the smoothed images.
Three of the sources lie too close to bright stars for
accurate photometry; we excluded these 
objects. We use the AB magnitude system with
$m_{AB}=-2.5\log f_\nu-48.60$, where $f_\nu$ is the flux of the 
source in units of erg~cm$^{-2}$~s$^{-1}$~Hz$^{-1}$.
The $1\sigma$ limits on the total magnitudes are 28.5 ($B$), 
28.4 ($V$), 28.2 ($R$), 27.3 ($I$), and 26.9 ($z'$)
based on measuring randomly positioned apertures
in the field. These are slightly more conservative (brighter) than 
those used in Barger et al.\ (2002), where the limits were 
computed for $3''$ diameter aperture magnitudes using the 
distribution of sky backgrounds in the individual pixels. The 
present limits deal better with correlated noise. 

%
%
\begin{inlinefigure}
\psfig{figure=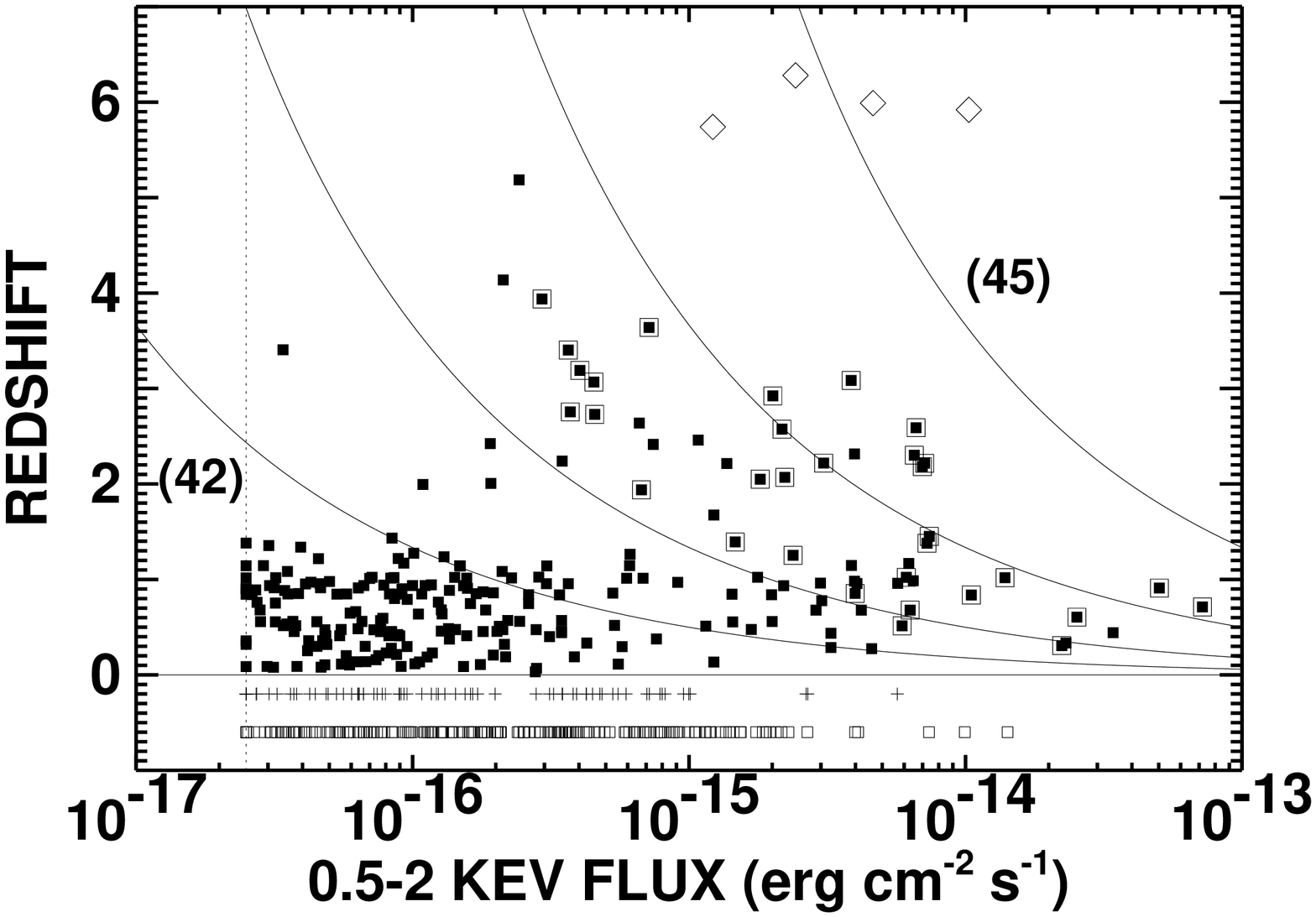,angle=90,width=3.5in}
\vspace{6pt}
\figurenum{1}
\caption{
Redshift versus $0.5-2$~keV flux for the entire
sample. Sources fainter than
$2.5\times 10^{-17}$~erg~cm$^{-2}$~s$^{-1}$
are shown at this limit. Three sources
with poor optical photometry are excluded from
the plot, as are the 13 spectroscopically identified
stars. Solid curves show flux versus redshift for
sources with a $2-8$~keV rest-frame luminosity
of $L_x=10^{42}$~erg~s$^{-1}$ (lowest curve),
$10^{43}$~erg~s$^{-1}$, $10^{44}$~erg~s$^{-1}$, and
$10^{45}$~erg~s$^{-1}$ (highest curve), computed with a
$K$-correction for an intrinsic $\Gamma=1.8$ power-law spectrum
where $\Gamma$ is the photon index.
Sources with broad-line optical spectra are
enclosed in a second, larger symbol. Spectroscopically
unidentified sources with $z'<25.2$ ($z'>25.2$) are
denoted by open squares (plus signs) at $z=-0.7$ ($z=-0.3$).
Large open diamonds denote the optically-selected
high-redshift quasars from the SDSS sample (Brandt et al.\ 2002;
Mathur et al.\ 2002).
\label{fig1}
}
\addtolength{\baselineskip}{10pt}
\end{inlinefigure}

Figure~\ref{fig1} shows redshift versus $0.5-2$~keV
flux. \markcite{barger02}Barger et al.\ (2002) and
A. J. Barger et al. (in preparation) give optical spectroscopic
identifications for 249 of the X-ray sources, including 16 off-axis
sources where the X-ray position is within the envelope
of a bright galaxy but not within $1.5''$ of the nucleus
(\markcite{horn02}Hornschemeier et al.\ 2002).
We analyze the colors for
the 423 (out of the 500 with accurate photometry) X-ray 
sources with optical counterparts brighter than the $5\sigma$ 
limit of $z'=25.2$. In Fig.~\ref{fig1}
we split the spectroscopically unidentified
counterparts into $z'<25.2$ (open squares 
at $z=-0.6$) and $z'>25.2$ (plus signs at $z=-0.3$).

The solid curves in Fig.~\ref{fig1} are for 
fixed rest-frame $2-8$~keV luminosities ranging from 
$10^{42}$~erg~s$^{-1}$ (lowest curve) to  
$10^{45}$~erg~s$^{-1}$ (highest curve).
The open diamonds show the $0.5-2$~keV fluxes of the 
highest redshift sources from the SDSS sample 
(\markcite{brandt02}Brandt et al.\ 2002; 
\markcite{mathur02}Mathur et al.\ 2002).
These form an optically complete sample
for $z>5.7$ and $z'<20$ over an area of 1500~deg$^2$
(\markcite{fan01b}Fan et al.\ 2001b). 
The present X-ray sample probes almost two orders of 
magnitude fainter than the typical X-ray flux seen in the 
SDSS sample; however, the area of sky covered is four orders of 
magnitude smaller. If the rest-frame optical to rest-frame 
X-ray flux ratios of high-redshift, low-luminosity 
sources are similar to those of the high-redshift, 
high-luminosity sources in the SDSS sample, then the magnitudes 
of the optical counterparts to the low-luminosity
sources would go as faint as $z'=25$, which is well-matched 
to our $z'=25.2$ color-selection limit. However, if 
instead high-redshift, low-luminosity sources have
lower rest-frame optical to X-ray flux ratios 
(\markcite{green95}Green et al.\ 1995; \markcite{vig03}Vignali,
Brandt, \& Schneider 2003), then the counterparts
could be fainter than $z'=25$.

\section{High Redshift AGN}
\label{hiz}

\subsection{Optically Bright Sample}
\label{z25}

High-redshift sources in the X-ray sample are easily
identified because the increasing line blanketing of the 
Lyman forest at high redshifts produces extremely 
large breaks across redshifted Ly$\alpha$ 
(e.g., \markcite{fan01b}Fan et al.\ 2001b). 
\markcite{songaila02}Songaila \& Cowie (2002) 
measured the average forest transmission 
in the redshift range $z=3.8-6.2$.
The magnitude break across Ly$\alpha$ is
\begin{equation}
\Delta{m} = 3.8 + 20.3\ {\rm log_{10}}\left({1+z}\over{7}\right) \,.
\label{eq:1}
\end{equation}
At $z=5$, where Ly$\alpha$ lies between 
$R$ and $I$, the break is 2.4 magnitudes, while at 
$z=6$, where Ly$\alpha$ lies between $I$ and $z'$, it is
3.8 magnitudes. If the epoch of reionization is at 
$z\sim 6.1$ (\markcite{becker01}Becker et al.\ 2001), then
the break is extremely large at redshifts higher than this.
The Ly$\alpha$ line moves through 
our longest wavelength ($z'$) band at 
$z\approx 6.5$, which is the largest redshift 
where a source would be seen in our optical data.

%
%
\begin{inlinefigure}
\psfig{figure=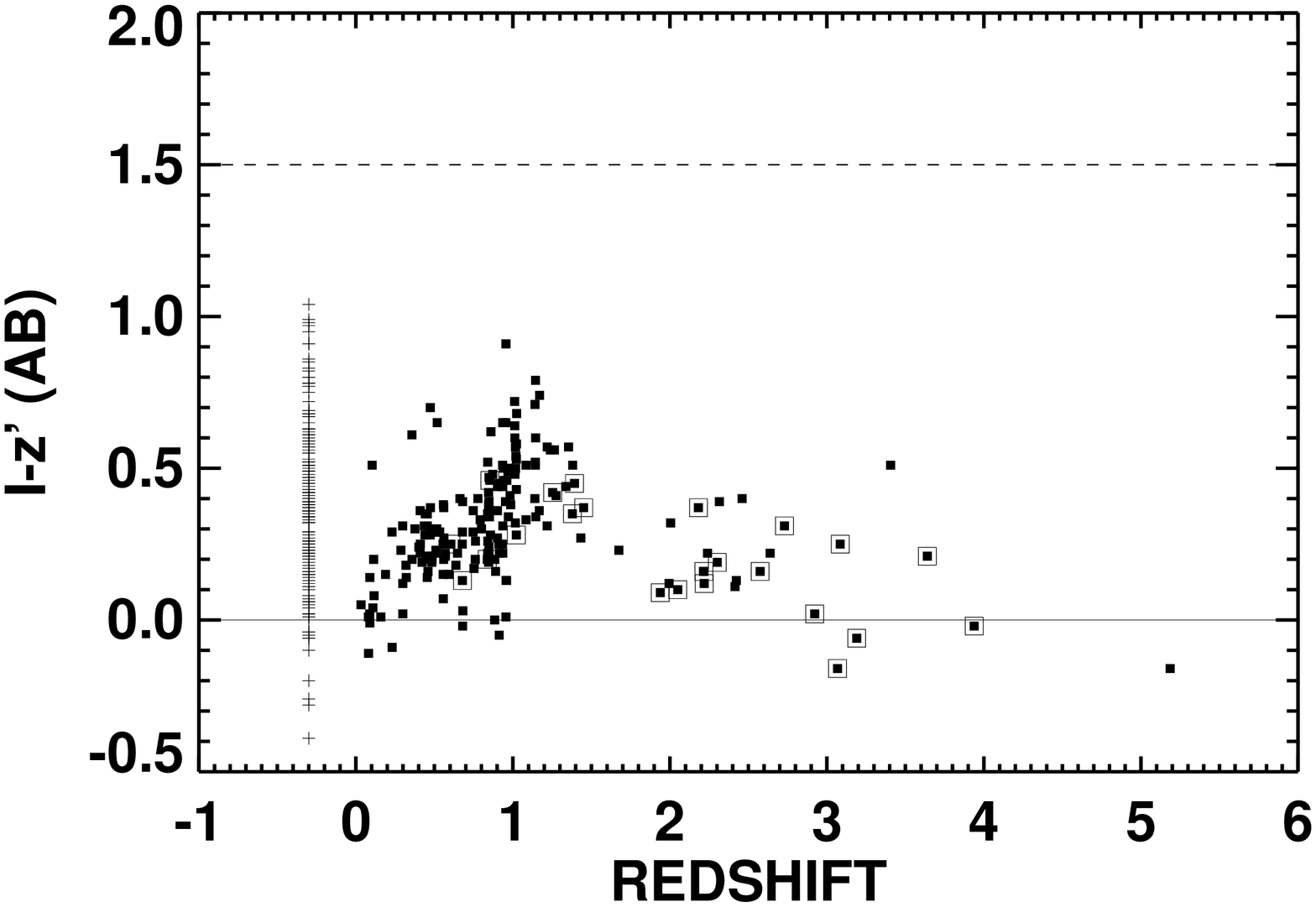,angle=90,width=3.5in}
\vspace{6pt}
\figurenum{2}
\caption{
$I-z'$ in the AB system versus redshift for the
423 sources with $z'<25.2$. Sources without
redshifts are shown at $z=-0.3$, and broad-line AGN are enclosed
in a second, larger symbol.
The dashed line ($I-z'=1.5$) is the color of $z\sim6$ sources
in the quasar and galaxy models described in the text.
None of the spectroscopically
unidentified sources are red enough in $I-z'$ to be at $z>6$.
\label{fig2}
}
\addtolength{\baselineskip}{10pt}
\end{inlinefigure}

Figure~\ref{fig2} shows $I-z'$ versus redshift (spectroscopically
unidentified sources are at $z=-0.3$) for our $z'<25.2$ sample.
We see from Fig.~\ref{fig2} that none of the sources lie at $z>6$,
since such sources would be redder than $I-z'>1.5$.
In Fig.~\ref{fig3} we show $V-I$ versus $I-z'$.  
The dashed line is the track with
redshift of the composite SDSS quasar spectrum of 
\markcite{vdb01}Vanden Berk et al.\ (2001), extrapolated with
a $f_\nu \sim \nu^{-0.46}$ power-law below Ly$\alpha$, modulated by 
the \markcite{songaila02}Songaila \& Cowie (2002)
forest transmissions, and convolved through the
Suprime-Cam filters. The extrapolation
below the Lyman continuum limit is not critical since
the high incidence of Lyman limit systems at these redshifts
truncates the spectrum at shorter wavelengths. The dot-dash
line shows the same calculation for a $f_\nu \sim \nu^{0}$ 
galaxy with no emission lines and a cut-off at $912$~\AA.
The deep forest absorption moves the 
tracks far from the galaxy populations. 
Apart from the one $z=5.19$ AGN (large
solid square) with $z'=23.9$, there are no other $z>5$ candidates.
Given the possibility that in a small number of cases we
may have chosen the wrong optical counterpart or the
colors are contaminated by a nearby object, we also
visually searched differenced images for objects
with red $V-I$ colors. No further candidates were found.

%
%
\begin{inlinefigure}
\psfig{figure=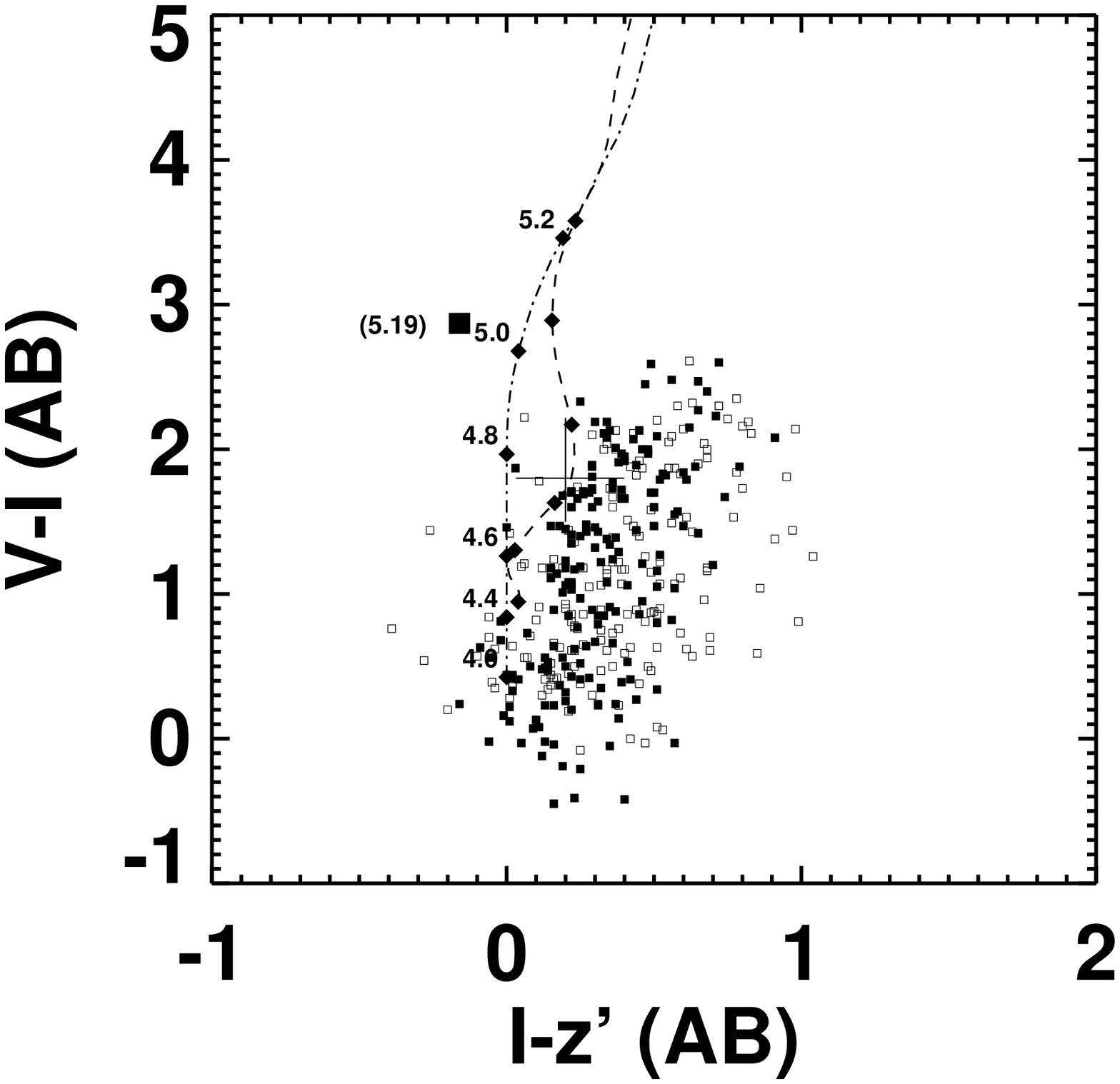,angle=90,width=3.4in}
\vspace{18pt}
\figurenum{3}
\caption{
$V-I$ versus $I-z^\prime$ for the $z^\prime<25.2$ sample.
Solid (open) squares denote galaxies with (without) redshift
identifications. All sources are detected above the $2\sigma$
level in all three bands. One sigma uncertainties are shown
for a $z'=25.2$ source lying at $I-z'=0.2$ and
$V-I=1.8$, typical of the location from which galaxies
might scatter into regions of the color-color space
and be misidentified as high redshift objects.
The one spectroscopically identified
$z=5.19$ source is shown with a large solid square.
The dashed (dot-dashed) curve shows the quasar (blue galaxy)
track with redshift.
Redshifts are given at the positions of
the diamonds on the tracks.
\label{fig3}
}
\addtolength{\baselineskip}{10pt}
\end{inlinefigure}

\subsection{Optically Faint Sample}
\label{optfaint}

Of the 500 X-ray sources in our sample, 77 
have $z'>25.2$. Two of these 
have spectroscopic redshifts ($z=3.40$ 
and $z=4.14$) from \markcite{barger02}Barger et al.\ (2002).
Of the remainder, 44 are detected above the $2\sigma$ 
level in $V$ or $B$ and cannot lie at $z>5$, leaving  
31 faint sources that could be at $z>5$.
Many of these faint sources 
may lie at lower redshifts (e.g., \markcite{yan02}Yan et al.\ 2002), 
and some may be false X-ray detections; about
half lie within a factor of 2 of the flux limit at their
off-axis radius, so this is an upper bound. 
Within a $6'$ off-axis radius where the sensitivity 
is relatively uniform, 9 of these optically faint
sources could have rest-frame $2-8$~keV
luminosities in excess of $10^{43}$ erg~s$^{-1}$, if they
were placed in the $z=5-6.5$ redshift range. Only 7 of 
these are not detected in $R$ or $I$ and could lie beyond $z=6$.
The cumulative surface density of field sources to $z'=25.2$ is
$\sim 1.2\times 10^{5}$~deg$^{-2}$, so we expect that about 30
of our sources with bright optical
identifications may be chance projections, with the
true optical counterparts being fainter. If we assume that 
these 30 sources are similar to the 77 isolated 
faint sources discussed above, then we can include these sources
in our candidate $z>5$ population with a scaling factor of 1.4. 

\markcite{haiman99}Haiman \& Loeb (1999) predicted a surface density
of $\sim50$ $z>5$ sources in a $6'$ radius circle
above a flux of $2\times 10^{-16}$~erg~cm$^{-2}$~s$^{-1}$ 
(observed $0.4-6$~keV). We find less than
four such $z>5$ objects, or a maximum of 5.6 
if we use the 1.4 scale factor discussed above.
The surface density is at least an order of
magnitude lower than the 
\markcite{haiman99}Haiman \& Loeb estimate,
strengthening the similar conclusion reached by 
\markcite{alex01}Alexander et al.\ (2001) and
\markcite{has02}Hasinger (2002) by about a factor of three.
 
\section{Discussion}
\label{secsdisc}

We compute the number density of $z=5-6.5$ sources in 
the rest-frame $2-8$~keV luminosity range $10^{43}$ to
$10^{44}$~erg~s$^{-1}$ following the procedures described 
in \markcite{cowie02b}Cowie et al.\ (2002b).
In Fig.~\ref{fig4}a we show the number density obtained
using only the one $z>5$ spectroscopic identification
(solid diamond). We also computed an upper limit to this
number density by assigning all the optically faint 
sources that could lie in the redshift interval redshifts at
the center of the interval; the horizontal bar corresponds to 
those with $L_x$ in the specified luminosity range.
In Fig.~\ref{fig4}a we compare our $z=5-6.5$ number density 
with the lower redshift number densities from 
\markcite{cowie02b}Cowie et al.\ (2002b) for both the $10^{43}$ 
to $10^{44}$~erg~s$^{-1}$ (solid circles) 
and $10^{44}$ to $10^{45}$~erg~s$^{-1}$ ranges (open circles). 
Again, the symbols denote the number densities obtained from only 
the spectroscopically identified AGN, while the horizontal bars
show the upper limits. The bars are not consistent with one another
because all the unidentified sources are included in each 
redshift bin if they lie in the specified luminosity range. 
To compare with the optically-selected samples 
of \markcite{boyle00}Boyle et al.\ (2000) and 
\markcite{fan01b}Fan et al.\ (2001b),
in Fig.~\ref{fig4}b we show the results for an 
$\Omega_M=1$, $\Omega_\Lambda=0$ 
cosmology with $H_o=50$~km~s$^{-1}$~Mpc$^{-1}$. 
The small number
uncertainties are large in Fig.~\ref{fig4}, but
the number densities in the $10^{43}$ to
$10^{44}$~erg~s$^{-1}$ range (AGN whose luminosities would
classify them as Seyferts) show a slow decrease with
increasing redshift from $z=0$ to $z=6$. This result holds
even within the systematic uncertainties shown by the
the horizontal bars. In contrast, the
number densities in the $10^{44}$ to $10^{45}$~erg~s$^{-1}$
range (AGN with quasar luminosities)
peak at $z=1.5-3$ and closely match in shape the evolution
of the optically-selected samples.
Figures~\ref{fig4}a and b show that this conclusion does
not depend on the cosmological geometry.

The number density at $z=5-6.5$ for
$L_x\approx 10^{43}$ to $10^{44}$~erg~s$^{-1}$
is about three orders of magnitude higher than the
optically-selected SDSS sample at
$L_x\approx 10^{45}$~erg~s$^{-1}$.
While this conclusion is based on the single $z>5$ object,
the fact that other high-redshift AGN have
also been found in the relatively small {\it Chandra} fields
(\markcite{silverman02}Silverman et al.\ 2002; $z=4.93$) and
in the small (74 arcmin$^2$) optical field of
\markcite{stern00}Stern et al.\ (2000; $z=5.5$) gives us
confidence that the CDF-N field is not anomalous. If we use
our spectroscopically identified point,
the slope $\beta$ of the LF ($\phi(L)dL=L^{-\beta}dL$)
would be about $2.6$, which is similar to the bright
end slope at lower redshifts in both the optical
(\markcite{pei95}Pei 1995)
and X-ray (\markcite{miyaji00}Miyaji et al.\ 2000).
If we include the unidentified objects, the maximum value of
$\beta$ would be about 3.
A detailed LF determination for this redshift interval will
require much larger area X-ray samples.

%
%
\begin{inlinefigure}
\psfig{figure=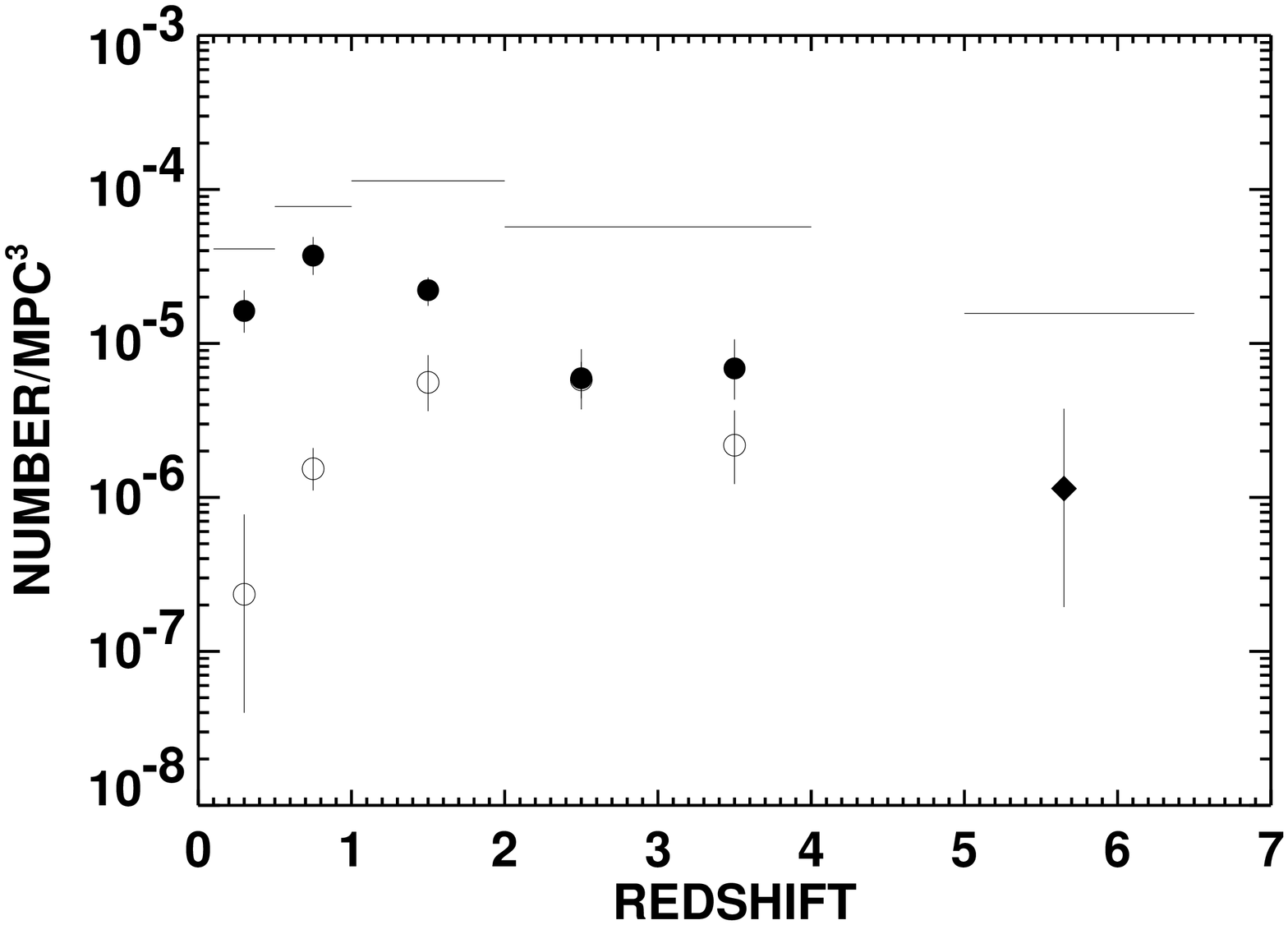,angle=90,width=3.2in}
\psfig{figure=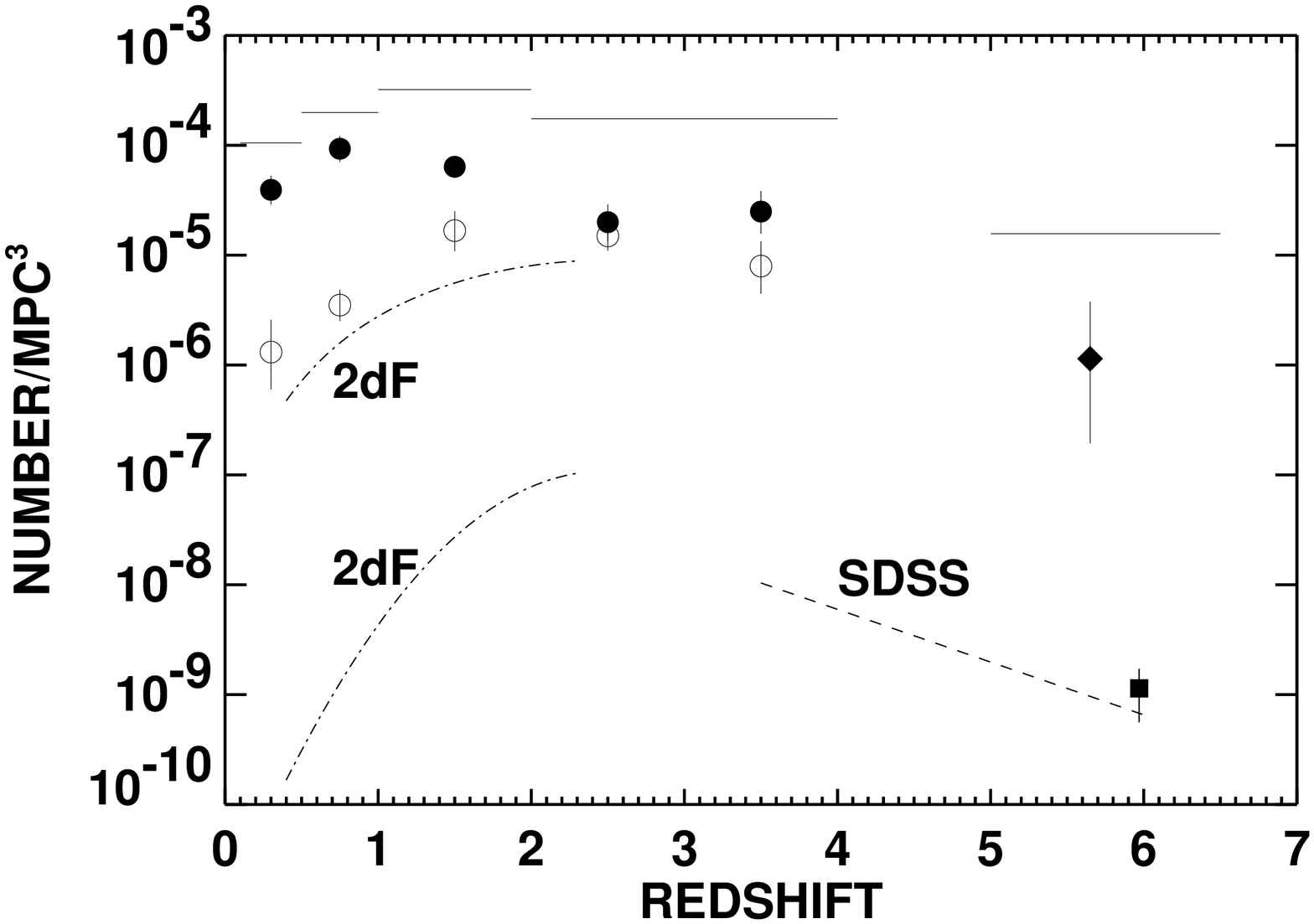,angle=90,width=3.2in}
\vspace{6pt}
\figurenum{4}
\caption{
(a)~Number density of sources with rest-frame $2-8$~keV
luminosities between $10^{43}$ and $10^{44}$~erg~s$^{-1}$
(solid symbols) and between $10^{44}$ and
$10^{45}$~erg~s$^{-1}$ (open symbols) versus redshift.
Diamond is from this paper, and circles are from Cowie et al.\ (2002b).
Points below (above) $z=2$ were determined from the observed-frame
$2-8$~keV ($0.5-2$~keV) sample. An intrinsic $\Gamma=1.8$ was assumed,
for which there is only a small differential $K$-correction
to correct to rest-frame $2-8$~keV. Poissonian $1\sigma$ uncertainties
are based on the number of sources in each redshift interval.
Horizontal bars show the maximal LF in the $10^{43}$ to
$10^{44}$~erg~s$^{-1}$ range found by assigning all the
sources that could lie in each redshift (and then luminosity)
interval a redshift at the center of the interval.
(b)~As in (a) but for an $\Omega_M=1$, $\Omega_\Lambda=0$
cosmology with $H_o=50$~km~s$^{-1}$~Mpc$^{-1}$.
Dot-dashed curves show the 2dF quasar LF of
Boyle et al.\ (2000): upper curve is for objects with
absolute $1450$~\AA\ magnitudes brighter than $-23$ (quasars)
that roughly matches our $L_x>10^{44}$~erg~s$^{-1}$ selection,
lower curve is for objects brighter than $-26.8$ that matches the SDSS
sensitivity to high-redshift quasars. Dashed line shows the SDSS
objects brighter than $-26.8$ in the $z=3.5$ to $z=6$ range.
Solid square and uncertainty is for the SDSS $z>5.7$ quasar sample
to $-26.8$ (Fan et al.\ 2001b).
\label{fig4}
}
\addtolength{\baselineskip}{10pt}
\end{inlinefigure}

\markcite{fan01b}Fan et al.\ (2001b) investigated the AGN ionizing
flux by fitting a variety of power-laws to extrapolate the bright 
SDSS data. A power-law 
with our observed slope of $\beta=2.6$ matched to the SDSS data 
fails by more than an order of magnitude to 
ionize the intergalactic medium at 
these high redshifts, even extrapolating to faint 
magnitudes (their Fig.~10). 
We may derive this result more directly 
by calculating the number of ionizing photons ($n_i$) per baryon 
($n_b$) produced in the redshift interval by the observed AGN.
We first sum the $K$-corrected fluxes of the sources
to determine the rest-frame $1450$~\AA\ background
light they produce and then convert this to
the number density of ionizing photons using
the form of the near-ultraviolet AGN spectrum given in 
\markcite{madau99}Madau, Haardt, \& Rees (1999), which gives
a ratio $\eta=4.7\times 10^{25}$ ionizing photons 
per erg~Hz$^{-1}$ at $1450$~\AA. ($\eta$ would be
less for a galaxy shaped spectrum.) We therefore have
\begin{equation}
{{n_i}\over {n_b}}={{4~\pi~\eta}\over{c~A~n_b}}
{\sum_{i} {f_i}} \,.
\label{eqnn}
\end{equation}
The $f_i$ are the fluxes corresponding to the $z'$ magnitudes 
for $z\sim 5.7$, and $A$ is the observed solid angle 
in steradians; we set $n_b=2.0\times~10^{-7}$ cm$^{-3}$. 
In the central $6'$ of the CDF-N field, which has the deepest 
uniform sensitivity and 
where the $z=5.19$ source is found, we obtain $n_i/n_b\approx 0.07$.
There is almost no change if we include all
the possible $z'>25.2$ sources in the region because these objects
are extremely faint 
in $z'$ and hence contribute little to the 
sum in Eq.~\ref{eqnn}. For a Poisson distribution
there is less than a 2\% probability of seeing 0 or 1 
sources when the true value is greater than 6, so at
$98\%$ confidence we can conclude that $n_i/n_b < 0.4$.
Thus, there are too few moderate luminosity AGN at $z=5-6.5$
to ionize the intergalactic medium.

\acknowledgements
Support came from the University of Wisconsin Research Committee 
(A.J.B.), the Alfred P. Sloan Foundation (A.J.B),
NSF grants AST-0084847 (A.J.B.), AST-0084816 (L.L.C.), and
AST-9983783 (D.M.A., W.N.B.), NASA grants DF1-2001X (L.L.C.) and
NAS 8-01128 (G.P.G.), and CXC grant G02-3187A (D.M.A., F.E.B., W.N.B.).

\end{document}